**A basal contribution from *p*-modes to the Alfvénic wave flux in the Sun's corona**

**Authors:** R. J. Morton[1], M. Weberg[1,2], J. A. McLaughlin[1]

[1]Department of Mathematics, Physics & Electrical Engineering, Northumbria University, Newcastle upon Tyne, NE1 8ST, UK

[2] National Research Council Research Associate residing at the Naval Research Laboratory, 4555 Overlook Avenue, SW Washington, DC 20375

**Many cool stars possess complex magnetic fields** [1] **that are considered to undertake a central role in the structuring and energising of their atmospheres** [2]**. Alfvénic waves are thought to make a critical contribution to energy transfer along these magnetic fields, with the potential to heat plasma and accelerate stellar winds** [3] [4] [5]**. Despite Alfvénic waves having been identified in the Sun's atmosphere, the nature of the basal wave energy flux is poorly understood. It is generally assumed that the associated Poynting flux is generated solely in the photosphere and propagates into the corona, typically through the continuous buffeting of magnetic fields by turbulent convective cells** [4] [6] [7]**. Here we provide evidence that the Sun's internal acoustic modes also contribute to the basal flux of Alfvénic waves, delivering a spatially ubiquitous input to the coronal energy balance that is sustained over the solar cycle. Alfvénic waves are thus a fundamental feature of the Sun's corona. Acknowledging that internal acoustic modes have a key role in injecting additional Poynting flux into the upper atmospheres of Sun-like stars has potentially significant consequences for the modelling of stellar coronae and winds.**

Alfvénic fluctuations have been observed regularly in the solar wind since the 1970's [8] [9] and are typically considered to be of solar origin. Their atmospheric counterpart was inferred from the non-thermal broadening of coronal emission lines [10], but only within the last decade have studies of the Sun's atmosphere been able to demonstrate unambiguously that magnetised plasma structures undergo displacements transverse to the magnetic axis [11] [12] [13]. Here we use infrared spectroscopic data (Fe xiii 1074.7nm emission line) taken from the Coronal Multi-channel Polarimeter (CoMP) coronagraph, which yield Doppler velocity time-series above the limb in the Sun's corona and provide a measure of Alfvénic wave motions along a viewer's line-of-sight [14] [15]. This is supported by extreme ultraviolet images of the corona from the 17.1nm (Fe ix) channel on-board the Solar Dynamics Observatory (SDO) Atmospheric Imaging Assembly (AIA), which enables direct measurement of the transverse oscillatory displacements of the corona's fine-scale magnetic structure [16] [17] (Fig. 1a, b, c). The data sets range between 2005 and 2015, hence covering various phases of the 11-year magnetic cycle, as the Sun's global magnetic geometry undergoes substantial changes (Supplementary Table 1). Since they can be subject to unambiguous and detailed measurements, we utilise the observed transverse displacements of magnetised plasma structures to probe the flux of Alfvénic wave energy.

The power spectral density (PSD) provides a means to investigate Alfvénic waves and is straightforward to obtain from CoMP Doppler velocity time-series (see Methods). The PSDs show evidence for power law behaviour and display an enhancement of power around 4 mHz, sitting atop of the power law base-line (Figure 1d). This behaviour has been noted in previous observations of individual coronal regions [11] [14]. Significantly, recent magnetohydrodynamic (MHD) wave models demonstrate the potential for coronal Alfvénic modes to be excited at the transition region [18] [19] [20], resulting from a double mode conversion of the Sun's internal acoustic (pressure or *p*-) modes that have leaked into the atmosphere through magneto-acoustic portals [21]. It is the observed enhancement of power that is considered to be the signature of Alfvénic waves generated by *p*-modes [11] [14] [20]. However, if *p*-modes are to play a crucial role in exciting coronal Alfvénic waves, their signature should have a spatially ubiquitous presence throughout the corona and over the solar cycle. Through examination of the Alfvénic waves associated with the power enhancement, we demonstrate that this is indeed the case.

The counterpart to the velocity fluctuations measured in CoMP is thought to be the swaying motions of coronal structures observed in SDO/AIA [13] [17], but no direct comparison has previously been undertaken. To remedy this, we measured large numbers of oscillatory Alfvénic waves in SDO/AIA (see Methods), where the imaging observations project the transient, transverse motions of plasma structures onto the plane-of-sky (Fig. 1b). The new detailed analysis of SDO/AIA data reveal that the coronal wave properties are more complex than previously thought. A bivariate relationship is found between frequency and amplitude (Fig. 1c), with the periods and amplitudes occupying greater ranges of values than previously reported [16, 17]. The increased statistics permit a way to cross-calibrate the two sets of waves observations, enabling an estimate for the time-averaged wave properties of a particular coronal region and, hence, the PSD of the oscillatory motions in SDO/AIA data (Fig. 1d).

The PSDs estimated from SDO/AIA data reveal the power spectra have a parabolic profile, which peaks around 3-4 mHz. Other peaks are also visible above the parabolic profile, but with the current uncertainties we cannot say whether these are genuine (See Methods). Comparison between the CoMP and SDO estimates for the PSDs of the Alfvénic waves reveals that the frequency location of enhanced power is congruous to the peak of the parabolic profile. Furthermore, the spectral indices from power law fits to both PSDs at frequencies > 4 mHz are in excellent agreement. The close relationship between PSDs suggests that the enhanced power in the CoMP data is due to the transient, oscillatory motions observed in SDO/AIA.

Given this relationship, the enhanced power in the CoMP PSDs then provides a distinct marker for oscillatory Alfvénic waves and enables us to examine their nature throughout the corona and solar cycle. Our analysis of the CoMP data reveals that the enhancement exists in a large majority (>95%) of coronal power spectra, meaning that Alfvénic waves are present throughout the entire corona. An example of the key measured parameters obtained for 10 May 2014 data are shown in Figure 2. We obtain similar findings when extending to data from different phases of the latest solar cycle (Fig. 3 & 4), from 2005, two years before solar minima, to the decline from maxima in 2015.

The frequency corresponding to the centre of the enhancement is found to fall within a narrow range, with its distribution having a mean and standard error of $4.0 \pm 0.1$ mHz and standard deviation of 1 mHz (Figs. 2 & 3a). While the sample populations are small for each yearly data set, a comparison of the distributions from different years suggests there is little variation in the centre values over the solar cycle. The enhanced power is distributed over a broader range of frequencies, and the characteristic width has a sharply peaked distribution around 0.12 frequency decades (Fig. 3b). The ubiquitous presence of the enhanced power through the corona is further evidenced in PSDs averaged across all the coronal time-series from a single day (Fig. 4), where it is seen that the signature of Alfvénic waves is still clear.

Aside from the enhanced power, the coronal PSDs from CoMP also display ubiquitous power-law-like behaviour, which indicates the presence of stochastic (or non-oscillatory) Alfvénic waves (Fig. 1d). However, it should be kept in mind we are examining a relatively short frequency range (0.1-10 Hz) and cannot determine if the coronal Alfvénic waves display scale-free behaviour. In spite of this shortcoming, we suggest the index from the fitting of a power law to the coronal PSDs can also provide insight into the nature of the coronal velocity fluctuations. This is based on previous analysis [13] that revealed the Alfvénic waves in coronal holes had similar spectral slopes to those found from studies of Alfvénic fluctuations in fast solar wind streams (where the measurements span a much greater frequency range) [9]. The distribution of power law indices shows a broad spread of values, with some evidence that it is bimodal (Fig. 3c). It is observed that the power law index typically decreases as the complexity of the local magnetic field increases (Fig. 2b), in line with previous results [14]. In particular, open field regions were found to have the shallowest slopes ($-1$) and active regions the steepest ($< -1.5$). The distributions from the different stages of the solar cycle show qualitatively similar shapes and spread, although there is some indication of variability (Fig. 3c).

Alfvénic waves that propagate in the corona are expected to interact non-linearly [4] [5] [6] and set up a turbulent MHD state [7] [22]. The spectral indices for the inertial range of a fully MHD turbulent state are thought to be $< -3/2$ (See Supplementary Text Section 1 for further discussion on the expected scaling). If MHD turbulence is in action throughout the Sun's low corona, the results would suggest that, in general, the currently observable frequency ranges typically only reveal the energy-containing (or correlative) scales. Observations of Alfvénic fluctuations in the fast solar wind demonstrate that the slope of power spectra is a function of frequency [23], with a change from the energy-containing scales (index $\approx -1$) to the inertial scales around 5 mHz at 0.3 AU. The frequency of this break-point decreases with distance from the Sun and is thought to be related to the decay of Alfvénic fluctuations with distance. With the presence of spectral indices close to $-1$ found in open and quiescent regions (Fig. 2 & [13] [14]), one may then naively expect such a break point to be located (if it exists) at higher frequencies in the lower corona. Hence, the current results imply the onset of inertial-scales in the corona occur generally at higher frequencies than those currently observable. However, the measured spectral indices suggest approximately 25% of the coronal velocity fluctuations have indices $< -3/2$, found primarily in active regions (Fig. 2), which could indicate the development of turbulence cascades at observable frequencies in more complex magnetic geometries.

Overall, the properties of the oscillatory Alfvénic waves (e.g., amplitude, frequencies) in the corona are unexpectedly similar throughout the solar cycle. Remarkably, the magnitude of the average coronal power enhancement varies only 10-15% between the dates analysed (Fig 4a). The near homogenous nature of the global wave properties occurs despite the fact that they show variability across different magnetic geometries (Fig. 2, e.g., quiescent Sun compared to active region). Moreover, at the peak of the enhancement, the extra power above the power law component is comparable to that from the stochastic contribution (Fig. 4b). Hence, the results imply that the Sun's internal acoustic oscillations play a significant role in exciting Alfvénic wave modes at this frequency range, and signify they provide a basal source of Poynting flux to the Sun's corona.

Recently there have been questions raised concerning the paradigm that a broad spectrum of coronal Alfvénic waves can originate from the photosphere. For example, the low ionisation fraction of the photosphere potentially restricts Alfvénic wave excitation [24] and causes additional wave damping from ion-neutral collisions [25]. Additionally, coupling of Alfvénic waves to slow magneto-acoustic waves via the pondermotive force can also lead to significant wave damping in the chromosphere [26], while the transition region provides a substantial reflective barrier for their propagation into the corona [6]. The excitation of coronal Alfvénic modes at the transition region via the mode coupling (beginning with the Sun's internal acoustic waves) circumvents these issues. The findings presented here reveal that, despite concerns, a broad spectrum of Alfvénic waves is present throughout the Sun's corona, and distributions of wave properties change minimally over the solar cycle (Figs. 3 & 4a).

Importantly, a basal flux of Alfvénic waves is a crucial requirement for any mechanism to be considered as a major component in the heating of quiescent and open field regions in the Sun's atmosphere, where the temperature and emission measure are relatively homogenous [27]. Using the wave measurements from SDO, an order of magnitude estimate for the observed Alfvénic wave energy flux in quiescent and coronal holes is 50-80 $Wm^{-2}$ (see Supplementary Text section 2), which falls below the standard values for coronal radiative losses (100-200 $Wm^{-2}$). However, the observed transverse motions are likely to be only a fractional detectable component of the total Alfvénic energy in the corona. Previous measurements of the transverse waves suggest a disparity between the observed wave amplitudes and those inferred from line widths [13] [28], hinting at the presence of additional Alfvénic modes in the corona that cannot currently be directly measured. It is expected that *p*-modes can excite a number of Alfvénic modes [20] and, if the non-thermal line widths are an accurate measure of the combined Alfvénic mode amplitudes [13] [28], this implies the energy requirements to counter radiative losses in the corona (excluding active regions) may easily be met by Alfvénic modes.

Furthermore, the findings may have significant implications for many Alfvénic wave models of solar plasma heating and wind acceleration [4] [5] [6], that largely neglect the additional energy contribution available from *p*-mode conversion in favour of excitation solely by the horizontal motions of the photospheric convection. Given that other cool, magnetised stars will have acoustic modes generated in their convective envelopes [29], *p*-modes potentially have an important role in energising other stellar coronae as well.

**Methods**

**Observations** The CoMP instrument [30] is able to make measurements of the Doppler shift of the Fe xiii emission line (1074.7 nm), giving line-of-sight averaged velocities for the coronal plasma. The data are processed by the CoMP data pipeline, correcting for a number of standard data artefacts, e.g., dark current, flat-fielding, subtraction of photospheric continuum emission. Final data products have a fixed cadence and spatial sampling of 4.46 arcsec. Specific details on cadence and length of time-series are given in Supplementary Table 1. The data is rigidly aligned using cross-correlation to remove frame-to-frame misalignments. Most data sets contain measurements that cover the entire off-limb corona (except the darkest coronal holes), except 2005 which only covers a quarter of the off-limb corona.

SDO/AIA [31] makes intensity measurements close to 17.1 nm using a relatively broad-band filter, of which a dominant contributor to emission in coronal plasma is Fe ix. The data is processed using the standard AIA data pipeline, with all data sets having a cadence of 12 s and a spatial sampling of 0.6 arcsec.

**Wave nomenclature** Here we use the term Alfvénic to refer to MHD wave modes that are highly incompressible, transverse, and for which the main restoring force is magnetic tension. Pure MHD modes only exist in idealised plasmas, i.e., systems with an ignorable/invariant coordinate. For example, a homogenous media can support Alfvén, slow & fast magneto-acoustic modes [20] and a cylindrical flux tube model with a boundary discontinuity can support an infinite number of modes, a kink mode being one example. However, the solar atmosphere is an inhomogeneous, continuous and highly dynamic plasma where MHD wave modes have a hybrid nature. This property is highlighted by adding a simple continuous radial density profile to the cylindrical model [16] [32] [33], where the kink mode becomes strongly coupled to a quasi-torsional Alfvén mode [32] and will transfer energy to this other mode as it propagates, meaning that wave identification as kink/Alfvén is not absolute and insufficient. While acknowledging that the observed transverse displacement of plasma structures observed in the CoMP and SDO/AIA data display characteristics of the idealised kink mode [15] – the term Alfvénic better acknowledges the rich and complex nature of the coupled wave system within a real continuous and inhomogeneous media.

**Automated wave detection** We have shown previously that the transverse modes in the corona can be directly measured with SDO/AIA even in regions with low signal-to-noise [17] [34]. These previous measurements were manual, meaning it required a substantial time to collect a statistically significant number of samples and the results were open to subjective event-selection biases. We now utilise an automated wave detection algorithm based upon the Fourier analysis of time-series (Fig. 1b), which permits the collection of large samples of wave measurements. Due to the nature of the Fourier analysis, the algorithm is tuned to preferentially detect oscillatory signals.

The automated detection of waves in SDO/AIA data relies on utilising time-distance plots to measure the plane-of-sky projection of transverse displacements of the fine-scale magnetic structure. We use the NUWT software [34], which finds and measures the location of structures in the time-distance plots, generating times-series for the location of each structure. The latest version of the software enables an automated analysis of large data volumes, using Fast Fourier Transforms to measure the frequency and displacement amplitude of the swaying motions.

We note that the SDO data possess an inherent limitation for measuring the waves due to the instrument resolution. Potentially this could alter the shape (and trend) of the bivariate distribution found in the measurements depicted in Fig 1c. The higher-frequency, oscillatory, transverse waves found typically have the smallest displacement amplitudes (although largest velocity amplitudes) and hence there is the potential that high-frequency, small-displacement amplitude waves are excluded from our measurements (i.e., waves occupying the lower right hand part of the frequency against velocity amplitude plot in Fig. 1c). However, we believe that this is not the case.

If we were to assume that there are unseen high-frequency waves with smaller displacement/velocity amplitudes, then this would lead to a smaller values of time-averaged power. Hence, the corresponding power spectral density will also be altered (Fig. 1d), and it would have smaller power at the higher frequencies than those currently shown. Such a shift in the high-frequency power would then lead to an inconsistency between the CoMP and SDO PSDs, which is not observed to be the case.

**Power spectral density**
*CoMP* – The power spectral densities (PSD) used in the following analysis are based on region-averaged PSDs, increasing the signal-to-noise over individual estimates of the PSD. The corona is divided into 5 degree sections, with each section containing between 130 and 200 individual time-series. The division starts at solar north. The choice of location of the section boundaries is arbitrary, in the sense the corona is equally divided up without care for magnetic geometry. We work under the assumption that in each section of the corona, the time-series are all different realisations of the same process. Certain sections of the corona are removed from analysis where there is no signal or only a very small number of time-series. The periodogram is a common way to display the time-averaged power as a function of frequency and is easily obtained from the discrete Fourier transform (DFT) of an evenly-sampled velocity time-series of length $N$ samples, evaluated at discrete frequencies, $f_j = j/N*dt$ for $j = 1,..,N/2$ (where $dt$ is the cadence).

It is known the power of the DFT in each frequency bin is distributed as $\chi$-squared with two degrees of freedom [35], hence, the mean log-power at each frequency ordinate is calculated in each section and a bias correction applied. The distribution of the mean log-power at each frequency ordinate is checked for normality by generating bootstrap samples (500) of the mean log-power from the sample distributions. The distributions of the bootstrap mean are compared at the 5% level to a Normal distribution via a Kolmogorov-Smirnov test (correcting for unknown mean and variance and also for multiple comparisons). It is found that the majority (>99.9%) of the bootstrap distributions show no evidence for differing from a Normal distribution at this level. From the distribution of bootstrap means, the standard deviation of the mean is measured for each frequency ordinate. A correction is applied to the standard deviations of the means to account for the dependence of time-series from neighbouring pixels in the data [36].

*SDO* – The calculation of the PSD for SDO/AIA data is different to that for CoMP. The time-series generated by NUWT are of varying length, so cannot be combined without correction. The Fourier power is directly comparable to mean-square velocity. Assuming each signal measured, $V$, is a sinusoid, it exists for a fixed time $T$ over the course of the observations period, $Ndt$, e.g.,

$$V = \begin{cases} A\sin(\omega t) & 0 < t \leq T \\ 0 & T < t \leq Ndt \end{cases} \quad (1)$$

The mean-squared velocity for each sinusoid is then:

$$\langle v_{\sin}^2 \rangle = \frac{1}{T} \int_0^T V^2 . \quad (2)$$

Over the total observation period the mean-squared velocity is:

$$\langle v_i^2 \rangle = \frac{T}{Ndt} \langle v_{\sin}^2 \rangle, \tag{3}$$

for each signal. For a fixed frequency, there may be a number, $P$, of independent signals measured over the total time, of differing length $T_1$, $T_2$ ,…. The total mean-square velocity is just an addition of the mean-square velocity of each signal. For simplicity, if assume each signal has a characteristic time T and amplitude, then the total mean-square velocity is given by:

$$\langle v_{tot}^2 \rangle = \frac{PT}{Ndt} \langle v_{\sin}^2 \rangle. \tag{4}$$

To calculate the actual total mean-square velocity, an estimate for the occurrence rate of the waves over the total observation period at a fixed frequency is required. To obtain this, we bin the measured signals as a function of frequency (i.e., the histogram in Fig. 1b), denoting the number of signals in each bin as $P$. Finally, each value is normalised by the total number of signals observed, $P_{tot}$, due to the time-distance diagrams spanning a large spatial extent. Mean-square velocities of signals that fall within each frequency bin are multiplied by the respective $P/P_{tot}$.

To obtain the average PSD from the individual values, non-parametric regression is performed using a Nadaraya-Watson Kernel estimator. This methodology utilises multivariate kernel density estimation as an alternative to histograms [37]. The kernel bandwidth is selected by cross-validation and bootstrapping is used to determine the standard errors (see Supplementary Text Section 3).

The lower number of measurements for waves with frequencies greater than ~4 mHz (Fig. 1c) means the significance of these peaks above the parabolic shape is small. We note that the SDO/AIA PSD is limited to a smaller range of frequencies than that from CoMP, which is mainly determined by the lifetimes of the features under observation (e.g., plumes, coronal loops). The visibility of the coronal features is potentially due to the thermodynamic cycle of the atmospheric plasma, which is thought to be subject to impulsive or episodic heating and cooling events [38]. This would impact upon the low-frequency part of the PSD (<0.8 mHz), potentially leading to an underestimate of the power here. Furthermore, Figure 1c demonstrates that the wave velocity amplitude increases with frequency, suggesting that the higher-frequency Alfvénic waves typically carry more energy than their lower-frequency counterparts. We do not believe this feature is an artefact of our methodology as testing of the sensitivity suggests we should be able to measure smaller disturbances. Moreover, the occurrence of high-frequency waves is sporadic in comparison, hence they contribute less energy on average, which is reflected in the PSD from both instruments. Any unresolved high-frequency waves with velocity amplitudes smaller than those shown would decrease the PSD at high frequencies and lead to disagreement with the CoMP data. The observed bivariate behaviour explains, in part, the approximate log-normal marginal distributions of the wave properties observed previously [17].

**Non-linear regression and model comparison** We perform non-linear regression on the mean log-power spectra in each section of the corona. Two models are used for regression to assess whether all power spectra studied display the enhancement that peaks around 4 mHz and to parametrise the properties of the enhancement. We start with the hypothesis that all power spectra display power-law behaviour, and some of these may have a power enhancement. The first model (M1) we fit to the data is a simple power law, namely:

$$P_{M1}(f_j) = A f_j^a + B \quad (5)$$

where $a$ is the power law index, $A$ is a constant of proportionality, and $B$ is a constant to describe the noise-dominated power ordinates at high frequencies (i.e., $B$ represents a Gaussian white noise process). The second model (M2) takes into account any excess power that causes a deviation from power law behaviour via the addition of an exponential term,

$$P_{M2}(f_j) = A f_j^a + B + C \exp\left(-\frac{(\ln f_j - D)^2}{2E^2}\right) \quad (6)$$

where constants $C$, $D$ and $E$ parametrise a log-normal function.

To fit the models to the data, we use maximum likelihood, maximising:

$$L = \prod_{j=1}^{N/2-1} (2\rho S_j^2)^{-1/2} \exp\left(-\frac{(P_j - P_M(f_j))^2}{S_j^2}\right) \quad (7)$$

for Normally-distributed data (neglecting the power at the Nyquist frequency), fitting the models in log-space (see Supplementary Text Section 3). Supplementary Table 2 provides the details of the parameters obtained from the fitting of the model given in Eq. 6.

In order to compare the model's ability to describe the power spectra, we utilise the Akaike Information Criteria (AIC). The AIC is based on information theory and provides a means to measure the information lost when fitting a model, enabling a comparison between goodness-of-fit and model complexity. The measure of the AIC is defined as:

$$AIC = 2k - 2\ln(L_{\max}) + \frac{2k(k+1)}{n-k-1}, \quad (9)$$

where $L_{max}$ is the maximum value of the likelihood surface, $k$ is the number of parameters and $n$ is the sample size. The model comparison is summarised by the quantity $\Delta_{AIC} = AIC_{M1} - AIC_{M2}$. The suitability of the $\Delta_{AIC}$ statistic for comparison between these two models is tested by performing Monte Carlo simulations. The simulations mimic the described analysis procedure, using time-series generated from the suggested models (see Supplementary Text Section 3) and run 5000 times. When using M2 as the underlying power model, the simulations demonstrate approximately 95% of the time model M2 is favoured correctly. Conversely, upon performing simulations using only a power law spectra (M1), it is found that the AIC test performs worse - identifying M1 power spectra only 70% of the time. The observed percentage of M2 type models preferred by the AIC test is in line with the expected rate of success from the Monte Carlo simulations, which would imply that all measured coronal power spectra contain a power enhancement. Moreover, if we are expecting to identify ~ 70% of the M1 type spectra from the AIC test, this further supports the idea that the potential number of M1 type power spectra in the corona is small.

**Data Availability** *The data that support the findings of this study are available from the corresponding author upon reasonable request. The SDO data are available from the Joint Science Operations Center - http://jsoc.stanford.edu. The CoMP data* *are available from the High Altitude Observatory data repository https://www2.hao.ucar.edu/mlso/mlso-home-page.*

**Corresponding Author** Please direct all correspondence and data requests to Richard J. Morton at richard.morton@northumbria.ac.uk

**Acknowledgements** All authors acknowledge this material is based upon work supported by the Air Force Office of Scientific Research, Air Force Material Command, USAF under Award No. FA9550-16-1-0032 and the Science & Technology Facilities Council via grant number ST/L006243/1. R.J.M. is grateful to the Leverhulme Trust for the award of an Early Career Fellowship and the High Altitude Observatory for financial assistance. M.J.W. acknowledges additional support from NASA grant NNH16AC39I and basic research funds from the Chief of Naval Research. R.J.M. is also grateful for discussions at ISSI Bern '*Towards Dynamic Solar Atmospheric Magneto-Seismology with New Generation Instrumentation*', and with G. Li and S. Tomczyk. The authors acknowledge the work of the NASA/SDO and AIA science teams, and National Center for Atmospheric Research/High Altitude Observatory CoMP instrument team.

**Author contributions** R.J.M. performed analysis of the CoMP data. R.J.M, M.W. and J.A.M. performed analysis of the SDO data. All authors discussed results and contributed to the writing of the manuscript.

**Competing interests** The authors declare no competing financial interests.

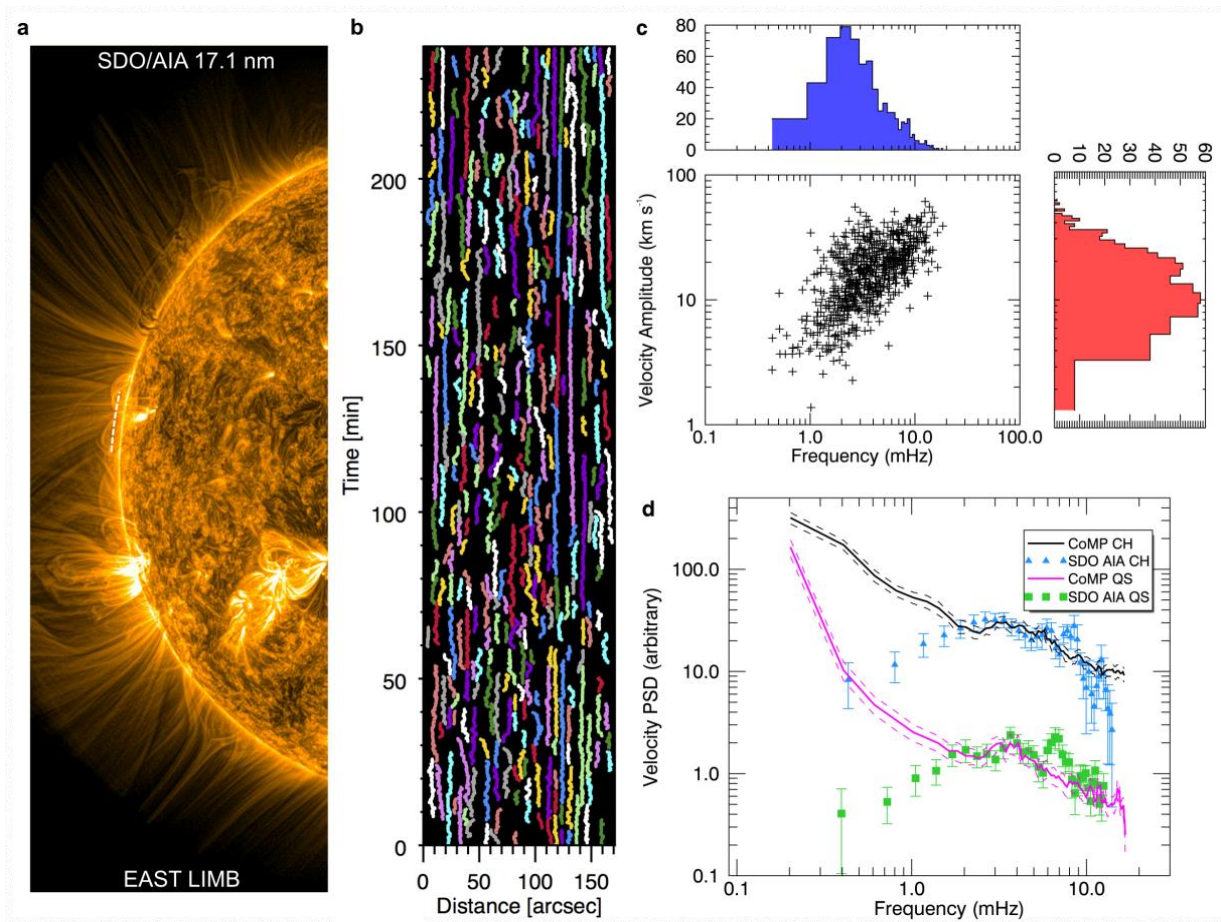

**Figure 1 The signature of coronal Alfvénic waves.** The corona is constituted of fine-scale magnetic structure as revealed by data from the SDO/AIA 17.1nm channel (**a**). Direct measurements of Alfvénic waves can be made through analysis of time-distance diagrams, with structures found to have displacements transverse to their axes. This is shown in **b**, where the coloured tracks mark oscillating structures in a time-distance diagram from the Quiet Sun (data taken from region shown by dashed slit in **a**). The measured properties of the waves are found to have a bivariate distribution (**c**), leading to approximately log-normal marginal distributions for both frequency (blue filled histogram) and amplitude (red filled histogram). Panel **d** shows the velocity power spectral density of Alfvénic waves from 27 March 2012 obtained from CoMP (solid lines; dashed lines are standard errors) and SDO/AIA (triangles and squares; error bars are standard errors). These are typical examples of average spectra found in an open magnetic field region (coronal hole - CH) and the quiescent Sun (QS). The PSD is shown on an arbitrary scale as CoMP is known to underestimate the power due to coarse spatial resolution. The QS data have also been offset by a factor of 1/10 for clarity.

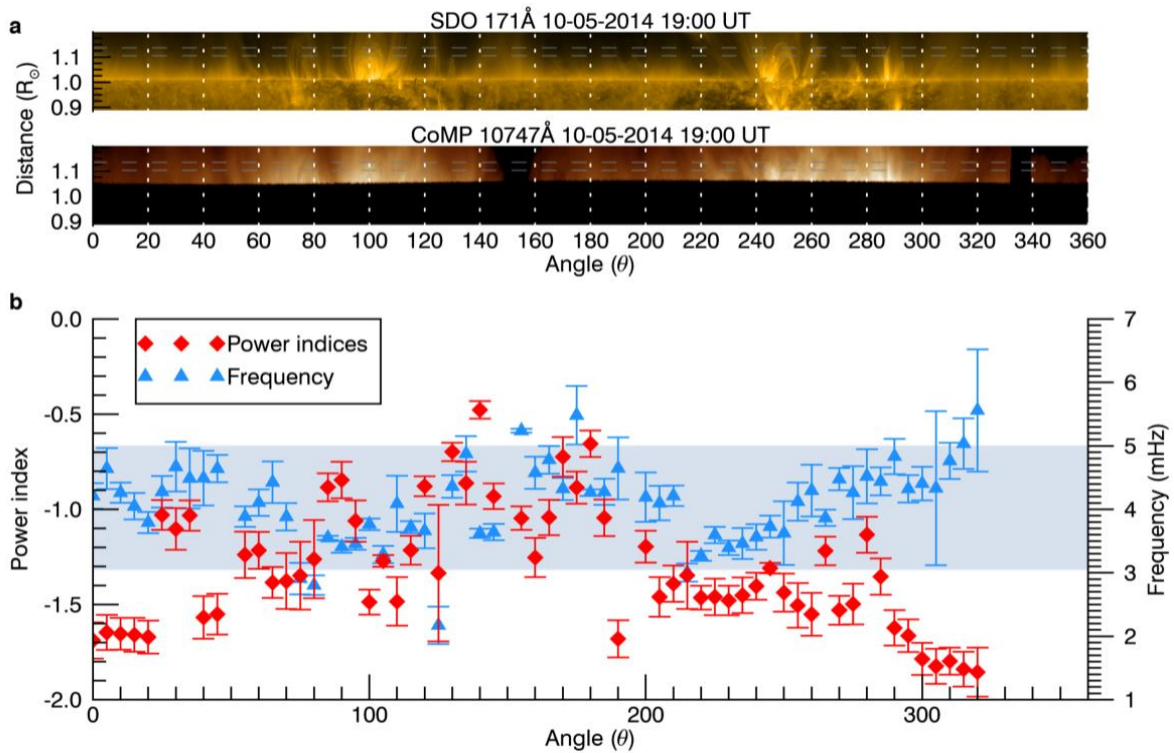

**Figure 2 Properties of Alfvénic waves throughout the corona.** The angle vs radius distance maps (**a**) show intensity images of the corona on 10 May 2014. The map starts at solar north and progresses clockwise around the limb. The corona was split into 5 degree wide segments and the CoMP data lying between the horizontal dashed grey lines was used in calculating region-averaged Doppler velocity power spectra in each segment. The active regions visible are located at $\theta \approx 70°$ (AR12050) $\theta \approx 100°$ (AR12049) and $\theta \approx 250°$ (AR12061), and a coronal hole is present at $\theta \approx 150°$. It is found that the majority of Doppler velocity power spectra follow a power law with enhanced power around 4 mHz. The values of spectral indices (red diamonds) and frequency values for the centre of enhanced power (blue triangles) are shown (**b**) from the non-linear regression to the power spectra, along with their standard error. The light blue shaded area highlights the 68% coverage of the central frequency Probability Density Function (shown in Figure 3a). The plot highlights the centre of enhanced power is constrained to a narrow range of frequencies, while the spectral index shows variation as the structure of the magnetic field changes between open and closed geometries.

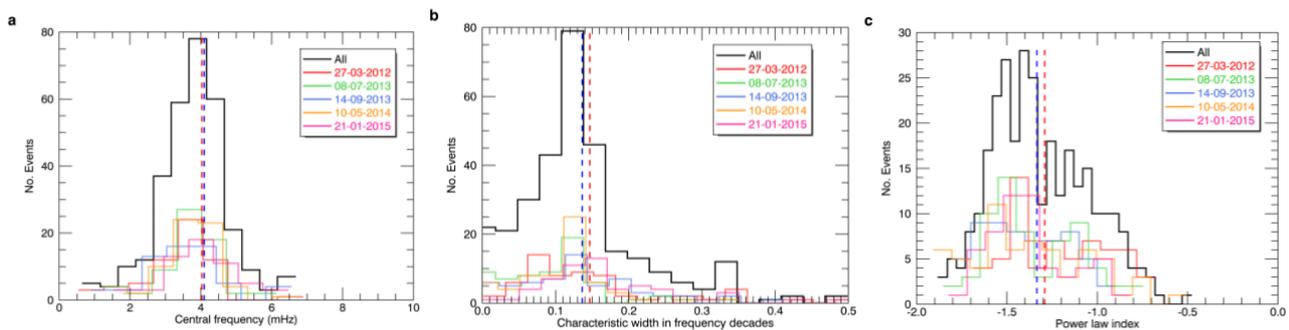

**Figure 3 Power spectra parameter distributions for Alfvénic waves.** The majority of coronal Doppler velocity power spectra are composed of a power law and a region of enhanced power (e.g., Fig. 1d). The characteristic parameters of these features are determined from region-averaged power spectra and are subject to non-linear regression. The frequency of the central location of the enhanced power shows a sharply peaked distribution **(a)** (central frequency of log-normal function fit to power spectra - parameter $D$ in Eq. 6 in Methods). The enhanced power is found to be present over a broader frequency range (**b**) as shown by the characteristic width of log-normal function. These distributions do not appear to vary significantly over the course of the solar cycle. The distribution of spectral indices from the coronal Doppler velocity power spectra **(c)** shows a greater variability between years, e.g., location of left-hand peak. The variability could be related to the changes in coronal magnetic geometry over the solar cycle, reflecting the variation in spectral index between different geometries indicated in Figure 2. However, we do not undertake further investigation here to confirm this due to the small yearly sample sizes. There is a total of 305 measurements from all dates (parameters for individual measurements given in Supplementary Table 2). The vertical red and blue dashed lines indicate the mean and median values, respectively.

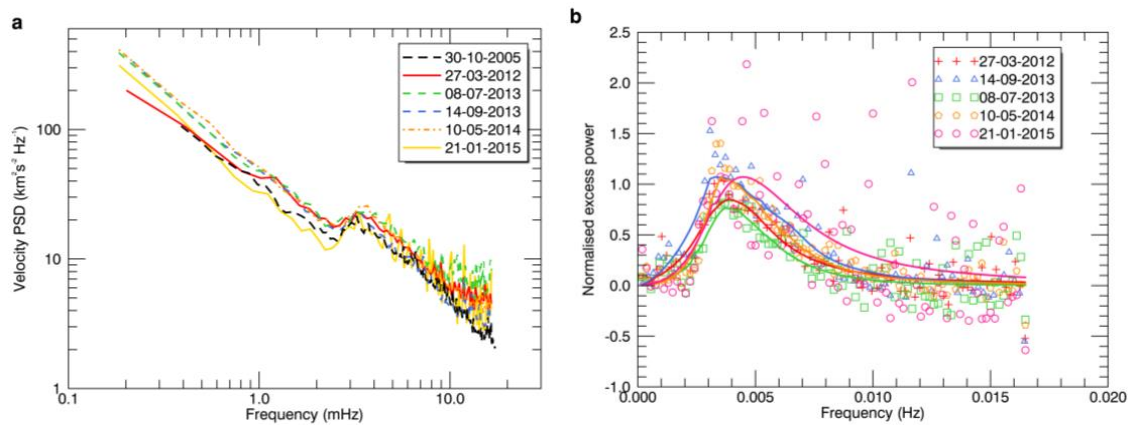

**Figure 4 Global measures of Alfvénic waves through the solar cycle.** The power spectra of velocity fluctuations after averaging over all eligible sections of the corona **(a)**. The dates chosen cover the solar minimum in 2005, close to the northern (late 2011) and southern (late 2013) sunspot number maxima and the declining phase of the cycle. The enhancement of power around 4 mHz is clear in the global averages and across the solar cycle. Upon calculating the fraction of power above that of the underlying power law **(b)**, the additional power at the peak of the enhancement is found to be comparable to the contribution from the stochastic velocity fluctuations. The data points are the average power spectra values and corresponding solid lines show the average values from the fitted models. Combined with the measured properties shown in Figure 3, the results reveal that the enhanced power is largely insensitive to the variations in global magnetic geometry of the corona over the Sun's magnetic activity cycle.

**A basal contribution from *p*-modes to the Alfvénic wave flux in the Sun's corona**

**Authors:** R. J. Morton[1]*, M. J. Weberg[1,2], J. A. McLaughlin[1]

[1]Department of Mathematics, Physics & Electrical Engineering, Northumbria University, Newcastle upon Tyne, NE1 8ST, UK

[2] National Research Council Research Associate residing at the Naval Research Laboratory, 4555 Overlook Avenue, SW Washington, DC 20375

*email to: richard.morton@northumbria.ac.uk

# Supplementary Material

**Supplementary Text**

**1. MHD Turbulence**
In incompressible turbulence with a strong mean magnetic field, the Alfvén effect is thought to be important and it is expected from phenomenological arguments that the inertial range should scale as -3/2 [1] [2] [3] [4]. However, it is currently unclear if this case is applicable to the low corona. The corona does contain a relatively strong mean magnetic field, i.e., the magnetic field strength is much greater than the perturbations, and it is clear from the observations that the fluctuations behave as Alfvénic waves. In such a case, it is not unreasonable to suggest that the system is essentially two dimensional and the Alfvén waves control the cascade rate, generating small-scales perpendicular to the mean magnetic field [3] [5].

However, if the strong mean magnetic field is not important, it expected that the scaling goes as -5/3 or steeper [3].

As the study of turbulence in the solar wind has a longer history, we may be able to draw parallels with results from observational studies. There are a number of reports of velocity fluctuations with spectral indices of -3/2 in the fast wind streams [4]. Although this is thought to be a transient state of the wind, with indications the spectra evolves to a Kolmogorov scaling (-5/3) as it propagates further from the Sun [6].

We note that the study of MHD Turbulence is an active and on-going area of research. Early phenomenological methods are now giving way to increasingly detailed computational methods. However, there is still on-going debate about the applicability of some of the phenomenological results, with simulations of MHD turbulence suggest an even steeper slope than the Kolmogorov spectrum (e.g., [3]).

**2. Alfvénic wave energy**
Here we provide further discussion on whether the energy content of the *p*-modes in the low solar atmosphere is enough to excite the observed wave modes.

We note that there is still an on-going debate over the amount of energy carried by transverse wave modes in the corona [7] [8] [9]. However, we provide a cautious estimate for the coronal energy flux of the observed transverse displacements in the quiescent regions and coronal holes. The energy flux, *F*, associated with the kink mode is given by [10], namely

$$F = f \rho v^2 c_p$$

where $f$ is the filling factor of over-dense plasma structures in the coronal region, $\rho=(\rho_i+\rho_e)/2$ is the mass density averaged over the coronal structure and ambient plasma, $v$ is the peak velocity amplitude, and $c_p$ is the propagation speed.

The values for the parameters used in the energy flux equation are given in Supplementary Table 3. The values of electron density, $n_e$, are taken from [11] and assumed to represent the average value of the coronal density. The mass density in a fully ionised plasma is then given by $\rho=\mu n_e$ where $\mu=2.12\times10^{-27}$ kg is the mean mass per particle. The value of $v_{obs}$ is the mean value of the measured velocity amplitudes from SDO. This value is the mean of the plane-of-sky motions, which are expected to contain oscillations not polarised in the plane of observation. There is no obvious reason that the wave modes would be polarised in a particular direction, hence we assume that the wave mode polarisation is uniformly distributed with respect to viewing angle. This implies the mean of the wave amplitudes will be reduced by a factor of $\sqrt{2}$ upon measurement (e.g., [12]). Hence we use $v = \sqrt{2}v_{obs}$. Finally, we note the values of energy flux given in Supplementary Table 3 use *f=1*, although the filling factor is likely less than 1. However, there is no consensus on what this value may be. The values of energy flux given in the main text assume a conservative value of *f=0.5*. The obtained values are in line with previous estimates of the coronal Alfvénic wave energy flux [8] [9].

The above calculation is only applicable to SDO/AIA data where the wave motions are resolved. The majority of the analysis in the main paper uses CoMP data and the amplitudes of the waves obtained from these measurements are under-resolved, making it difficult to convert the measured velocity values to actual amplitudes. It is envisioned that further combined analysis with SDO will help to resolve this issue.

The magneto-acoustic wave energy flux in quiet Sun magnetic elements at 400 km above the photosphere is found to be on the order of ~3000 W/m$^2$ at 3 mHz [13], with the wave energy integrated over a wider frequency range likely to be larger than this. If we can assume the magneto-acoustic waves at 400 km have not yet passed through the equipartition layer, then only 1-7% of this flux needs to be converted to meet the requirements for exciting the observed Alfvénic waves. Given that the current theoretical estimates suggest up to 30% of the *p*-mode energy can be converted to Alfvénic modes [14], the measured energy flux of magneto-acoustic energy in the lower solar atmosphere appears not to preclude the *p*-mode conversion model.

### 3. Further methodology and software details

The *np* package in *R* is used to perform the Kernel Density Estimation (Methods – Power spectral density) [15].

To undertake the non-linear regression (Methods - Non-linear regression and model comparison) we use the *mpfit* code in IDL [16].

To generate time-series in the Monte-Carlo simulations we use an inverse Fourier Transform method [17].

**Supplementary Tables**

| Date | Cadence (s) | Duration (s) | Number of coronal regions |
|---|---|---|---|
| 30-10-2005 | 29 | 10179 | - |
| 27-03-2012 | 30 | 4,890 | 61 |
| 08-07-2013 | 30 | 4,620 | 64 |
| 14-09-2013 | 30 | 5,460 | 60 |
| 10-05-2014 | 30 | 5,460 | 64 |
| 21-01-2015 | 30 | 5,460 | 56 |

**Supplementary Table 1 CoMP data specifics** Properties of the CoMP data sets used.

| Date | Angle | A ($10^{-6}$ km$^2$ s$^{-1}$) | $\alpha$ | B ($10^{-3}$ km$^2$ s$^{-1}$) | C ($10^{-3}$ km$^2$ s$^{-1}$) | D ln(Hz) | E ln(Hz) | $\chi^2_\nu$ |
|---|---|---|---|---|---|---|---|---|
| 2012-03-27 | 0.  | 0.16±0.16     | -1.60±0.14 | 1.63±0.11 | 3.97±0.47 | -5.57±0.07 | 0.45±0.06 | 2.76 |
| 2012-03-27 | 5.  | 1.35±1.24     | -1.27±0.13 | 1.96±0.15 | 4.07±0.50 | -5.46±0.04 | 0.32±0.04 | 2.23 |
| 2012-03-27 | 10. | 43.38±45.85   | -0.75±0.14 | 0.84±0.49 | 2.55±0.43 | -5.45±0.07 | 0.37±0.07 | 1.41 |
| 2012-03-27 | 15. | 27.89±18.46   | -0.88±0.09 | 0.21±0.34 | 1.73±0.41 | -5.29±0.06 | 0.23±0.06 | 1.51 |
| 2012-03-27 | 20. | 59.64±44.12   | -0.78±0.10 | 0.07±0.54 | 2.05±0.46 | -5.30±0.06 | 0.24±0.06 | 1.14 |
| 2012-03-27 | 25. | 7.10±3.47     | -1.14±0.07 | 0.95±0.21 | 5.59±1.57 | -5.51±0.01 | 0.04±0.01 | 1.33 |
| 2012-03-27 | 30. | 0.35±0.32     | -1.53±0.13 | 1.08±0.11 | 1.84±0.37 | -5.35±0.08 | 0.36±0.07 | 1.16 |
| 2012-03-27 | 35. | 0.21±0.31     | -1.57±0.19 | 0.76±0.17 | 0.98±0.64 | -5.61±0.47 | 0.64±0.37 | 1.66 |
| 2012-03-27 | 40. | 0.26±0.48     | -1.46±0.25 | 0.91±0.10 | 1.81±0.66 | -5.84±0.24 | 0.60±0.17 | 2.61 |
| 2012-03-27 | 45. | 0.34±0.31     | -1.49±0.13 | 0.98±0.09 | 1.66±0.30 | -5.41±0.08 | 0.35±0.07 | 2.77 |
| 2012-03-27 | 50. | 0.21±0.31     | -1.48±0.19 | 0.89±0.08 | 2.62±0.39 | -5.66±0.09 | 0.46±0.07 | 3.70 |
| 2012-03-27 | 55. | 18.77±13.69   | -0.86±0.10 | 0.01±0.23 | 1.67±0.38 | -5.66±0.12 | 0.41±0.10 | 2.62 |
| 2012-03-27 | 60. | 0.44±0.46     | -1.39±0.14 | 0.46±0.08 | 2.47±0.34 | -5.53±0.06 | 0.35±0.05 | 3.00 |

**Supplementary Table 2 Results from enhanced power model fit** Excerpt from extended table detailing all parameters and 1 sigma uncertainties from non-linear least squares fitting. Extended table available in machine readable format as an additional supplementary file.

|  | Coronal Hole | Quiet Sun |
|---|---|---|
| $n_e$ ($log_{10}$ cm$^{-3}$) | 8.3 ± 0.1 | 8.5 ± 0.1 |
| $v_{obs}$ (km s$^{-1}$) | 18.5 ± 0.4 | 16.5 ± 0.4 |
| $c_p$ (km s$^{-1}$) | 420 ± 80 | 420 ± 120 |
| $F(f = 1)$ (W m$^{-2}$) | 108 ± 46 | 154 ± 57 |

**Supplementary Table 3 Wave and plasma parameters.** The parameters given are used in the calculation of coronal Alfvénic wave energy flux.

**Supplementary references**